# Revisiting the Z-dependence of the electron density at the nuclei

Alireza Marefat Khah and Shant Shahbazian[*]

Faculty of Chemistry, Shahid Beheshti University, G. C., Evin, Tehran, Iran
19839. P.O. box 19395-4716
Tel/Fax: 98-21-22431661

E-mail:

Shant Shahbazian: chemist_shant@yahoo.com

[*] Corresponding author

**Abstract**

A new formula that relates the electron density at the nucleus of atoms, $\rho(0,Z)$, and the atomic number, $Z$, is proposed. This formula, $\rho(0,Z) = a(Z - b\sqrt{Z})^3$, contains two unknown parameters ($a,b$) that are derived using a least square regression to the ab initio derived $\rho(0,Z)$ of Koga's dataset from *He* ($Z=2$) to *Lr* ($Z=103$) atoms (Theor Chim Acta 95, 113 (1997)). In comparison to the well-known formula, $\rho(0,Z) = aZ^b$, used for the same purpose previously, the resulting new formula is capable of reproducing the ab initio $\rho(0,Z)$ dataset an order of magnitude more precisely without introducing more regression parameters. This new formula may be used to transform the equations that relate correlation energy of atoms and $\rho(0,Z)$ into simpler equations just containing the atomic number as a fundamental property of atoms.

Some years ago Liu and Parr proposed an empirical relationship between the electron correlation energies of atoms ($E_{corr}$) and the electron density on the nuclei $\rho(0,Z)$; $E_{corr} = CN\rho(0,Z)Z^{-\gamma}$ ($C$ and $\gamma$ are constants while $Z$ and $N$ are the atomic number and the number of electrons, respectively) [1]. Using a relatively large dataset of the correlation energies and $\rho(0,Z)$, employing a least square regression, they succeeded to determine the optimum values for the constants. The resulting compact equation is capable of reproducing atomic correlation energies accurately, albeit in a narrow range of $Z$. Interestingly $\rho(0,Z)$ appears as an "independent" variable in their equation that is used besides $N$ and $Z$ as fundamental parameters of an atom. Because of the importance of the correlation energies, it is tempting to trace a theoretical route to this equation and the origin of the significance of $\rho(0,Z)$ as seemingly something unrelated, at least directly, to the correlation energy. An alternative possibility is that $\rho(0,Z)$ itself is related in a compact way to $Z$ and its role in the aforementioned equation is just "absorbing" part of the Z-dependence of the atomic correlation energies. If true, then the equation may be transformed into a new equation just depending on the fundamental parameters ($N$ and $Z$), i.e. $E_{corr}(N,Z)$. Accordingly, in this short communication the $Z$-dependence of $\rho(0,Z)$ is considered.

The direct $Z$-dependence of $\rho(0,Z)$ as well as the indirect $Z$-dependence, through considering the relationship of $\rho(0,Z)$ with various quantum observables, were considered in several previous theoretical and computational studies [2-16]. Seeking the relationship of $\rho(0,Z)$ with various quantum observables was particularly stimulated from the derivation of lower and upper bounds for $\rho(0,Z)$ [3], through a set inequalities, which triggered several subsequent studies [6-10]. However, comparison of the computed bounds for large sets of atoms, using inequalities containing various moments of the position, i.e. $\langle r^k \rangle$, with ab initio $\rho(0,Z)$ derived directly from atomic Hartree-Fock wavefunctions demonstrated just a semi-quantitative agreement [8,9]. As realized early by Cioslowski [8], from a theoretical viewpoint by extending the original inequalities it is possible to set more accurate lower/upper bounds for $\rho(0,Z)$. However, this extension is not computationally cost effective since one must include higher moments of position in inequalities and compute them with a high precision. As an alternative and to simplify the whole analysis, the Bare Coulomb Model (BCM), which neglects the electron-electron interactions in atom but retains electron-nucleus interactions, has been employed as a simple yet analytically tractable model for real atoms. Although this simplified model is capable of explaining certain aspects of the $Z$-dependence of $\rho(0,Z)$ in real atoms, it is unable to reproduce the ab initio $\rho(0,Z)$ datasets quantitatively [12].

From a semi-empirical viewpoint one may use theoretically motivated direct $Z$-dependent equations and then supplement them with appropriate empirical parameters to be determined from a regression to a set of known $\rho(0,Z)$ dataset. This path has been pursued using a simple compact equation:

$$\rho(0,Z) = aZ^b \tag{1}$$

where $a$ and $b$ are determined by a least square regression ($a \approx 3$, $b \approx 0.5$) [5]. The contribution of 1s orbital to $\rho(0,Z)$ derived assuming a single s-type Slater function, $\rho_{1S}(0,Z) = (2/\pi)Z^3$, has been the theoretical motivation behind this equation [5] (see the supporting information for details). Although the resulting optimized equation is superior in reproduction of the ab initio computed $\rho(0,Z)$ datasets in comparison to the previously mentioned inequalities, it is yet incapable of an accurate reconstruction of the Hartree-Fock computed $\rho(0,Z)$.

Based on these reports it was decided to maintain the above mentioned semi-empirical viewpoint but seek for a more accurate but yet compact equation. At first stage equation (1) was used in a regression procedure to reproduce Koga's ab initio $\rho(0,Z)$ dataset derived at the Hartree-Fock level and depicted in Figure 1 ($Z = 2-103$) [17] (see the supporting information for the dataset). The resulting optimized parameters, gathered in Table 1, are not far from what was reported previously [5] while the detailed results for each atom have been offered in the supporting information. The mean absolute error is not large, however the errors are not evenly distributed among atoms; the equation works well for large $Z$ and the percentage error is less than 1% for $Z > 31$ and less than 0.1% for $Z > 53$, but for smaller atomic numbers the percentage error grows rapidly. While equation (1) describes the main trend of the $Z$-dependence correctly, it is not capable of reproducing the ab initio $\rho(0,Z)$ dataset with reasonable accuracy, a percentage error less than 1%, and for a more quantitate description one must go beyond equation (1).

After some heuristic attempts the following equation was proposed based on modification of the BCM model (see the supporting information for details):

$$\rho(0,Z) = a\left(Z - b\sqrt{Z}\right)^3 \tag{2}$$

This equation is slightly more complicated than equation (1) but still has just two unknown parameters to be determined in a least square regression procedure. Table 1 offers the optimized parameters as well as the results of statistical analysis while the detailed results for each atom have been gathered in the supporting information. Evidently, in comparison to the results of equation (1), equation (2) is a mark refinement and now only for four atoms ($Z = 2, 4-6$) the percentage error is more than 1%; even in these cases the maximum percentage error does not exceed from 3.5% while for $Z > 18$ the percentage error is equal or less than 0.1%. By the way, like equation (1), the errors are not evenly distributed among atoms and no simple $Z$-dependent pattern emerges for the remaining absolute errors making further modification of equation (2) practically hard. Since the number of parameters is equal in equations (1) and (2) the much better performance of equation (2) does not seem to originate from its more "flexibility". It is tempting to introduce an effective atomic number, $Z_{eff} = Z - b\sqrt{Z}$, and conceiving $b\sqrt{Z}$ as some kind of average screening constant that transforms equation (2) to: $\rho(0,Z) = a(Z_{eff})^3$. However, at present state of knowledge no theoretical reasoning justifies this interpretation.

In a nutshell quantitative reproduction of the $Z$-dependence of the ab initio electron densities at the nucleus is feasible without any need to complicated formulas. This cast some doubt that $\rho(0,Z)$ may be treated as an independent variable in equations used to reproduce atomic correlation energies and probably by adding an extra non-linear parameter to such equations, e.g. $b$ in the case of equation (2), one may transform them into formulas just based on the fundamental parameters of atom ($N$ and $Z$). A detailed analysis in this direction will be offered in a separate forthcoming report.

## Acknowledgments

The authors are grateful to Prof. Toshikatsu Koga for sharing his database of atomic electron densities and to Cina Foroutan-Nejad for reading a previous draft of this paper and his helpful suggestions. This communication is part of the results of the project "3D structure of free atoms" done during the period 2010-2012 and supported by the research council of Shahid Beheshti University.

Table 1- The optimized parameters and some statistical parameters resulting from the least square regression of equations (1) and (2) to the dataset (All results are given in atomic units).

| Equation | Optimized parameters | | Statistical parameters | | |
|---|---|---|---|---|---|
| | $a$ | $b$ | $MAE^*$ | $MaxAE^{**}$ | $MaxPE^{***}$ |
| (1) | 0.605959 | 3.038611 | 115.3 | 326.0 ($Z=103$) | 38.5% ($Z=2$) |
| (2) | 0.773780 | 0.220476 | 22.0 | 64.8 ($Z=79$) | 3.5% ($Z=2$) |

* The mean absolute error (MAE) was computed according to: $\sum_{Z=2}^{103}\left(\frac{|\rho(0,Z)-\rho^{ab\ initio}(0,Z)|}{102}\right)$.

** Maximum absolute error (MaxAE): $Max|\rho(0,Z)-\rho^{ab\ initio}(0,Z)|$.

*** The Maximum percentage error (MaxPE) was computed according to: $\left(|\rho(0,Z)-\rho^{ab\ initio}(0,Z)|\big/\rho^{ab\ initio}(0,Z)\right)\times 100$

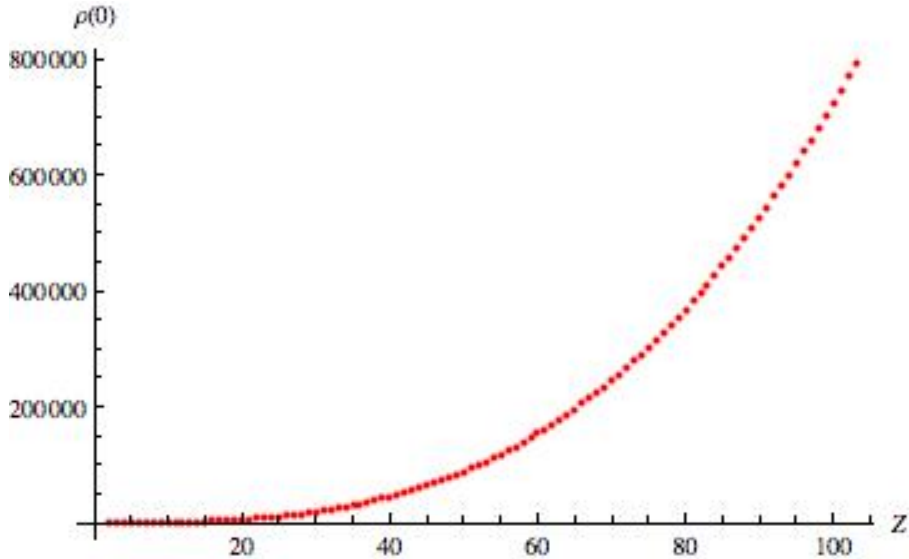

Figure 1- The ab initio $\rho(0,Z)$ (in atomic units) against the atomic number ($Z=2-103$).

# Supporting Information

Revisiting the Z-dependence of the electron density at the nuclei

Alireza Marefat Khah and Shant Shahbazian[*]

Faculty of Chemistry, Shahid Beheshti University, G. C., Evin, Tehran, Iran
19839. P.O. box 19395-4716
Tel/Fax: 98-21-22431661

E-mail:

Shant Shahbazian: chemist_shant@yahoo.com

[*] Corresponding author

# Table of contents

The ab initio and computed $\rho(0,Z)$ based on equation: $\rho(0,Z) = aZ^b$ ($a \approx 0.605959$, $b \approx 3.038611$) and the absolute error, $\rho(0,Z) - \rho^{ab\,initio}(0,Z)$, and the percentage errors $\left(\left|\rho(0,Z) - \rho^{ab\,initio}(0,Z)\right| / \rho^{ab\,initio}(0,Z)\right) \times 100$. All results are in atomic units ($Z = 2 - 103$).

| Atomic number | Ab initio | Predicted | Absolute error | Percentage error |
|---|---|---|---|---|
| 2 | 3.5959183 | 4.9791652 | 1.38 | 38.47 |
| 3 | 13.8148199 | 17.0698402 | 3.26 | 23.56 |
| 4 | 35.3877167 | 40.9137919 | 5.53 | 15.62 |
| 5 | 71.9213707 | 80.6012187 | 8.68 | 12.07 |
| 6 | 127.4579724 | 140.2628471 | 12.80 | 10.05 |
| 7 | 205.9683473 | 224.0618577 | 18.09 | 8.78 |
| 8 | 311.6608926 | 336.1885570 | 24.53 | 7.87 |
| 9 | 448.3222768 | 480.8565853 | 32.53 | 7.26 |
| 10 | 619.9220741 | 662.3000736 | 42.38 | 6.84 |
| 11 | 833.7575330 | 884.7714308 | 51.01 | 6.12 |
| 12 | 1093.7178333 | 1152.5395705 | 58.82 | 5.38 |
| 13 | 1402.8456067 | 1469.8884576 | 67.04 | 4.78 |
| 14 | 1765.6069489 | 1841.1158947 | 75.51 | 4.28 |
| 15 | 2186.3141472 | 2270.5324929 | 84.22 | 3.85 |
| 16 | 2669.4696949 | 2762.4607881 | 92.99 | 3.48 |
| 17 | 3219.1903560 | 3321.2344732 | 102.04 | 3.17 |
| 18 | 3839.7818294 | 3951.1977251 | 111.42 | 2.90 |
| 19 | 4538.6539527 | 4656.7046099 | 118.05 | 2.60 |
| 20 | 5319.6071378 | 5442.1185530 | 122.51 | 2.30 |
| 21 | 6182.3043131 | 6311.8118639 | 129.51 | 2.09 |
| 22 | 7133.2302762 | 7270.1653082 | 136.94 | 1.92 |
| 23 | 8176.9553727 | 8321.5677196 | 144.61 | 1.77 |
| 24 | 9313.7337083 | 9470.4156463 | 156.68 | 1.68 |
| 25 | 10560.0774220 | 10721.1130282 | 161.04 | 1.52 |
| 26 | 11908.7041093 | 12078.0708997 | 169.37 | 1.42 |
| 27 | 13367.3818433 | 13545.7071169 | 178.33 | 1.33 |
| 28 | 14940.5685578 | 15128.4461046 | 187.88 | 1.26 |
| 29 | 16625.2578185 | 16830.7186220 | 205.46 | 1.24 |
| 30 | 18447.6758062 | 18656.9615453 | 209.29 | 1.13 |
| 31 | 20397.2023524 | 20611.6176639 | 214.42 | 1.05 |
| 32 | 22480.1179151 | 22699.1354912 | 219.02 | 0.97 |
| 33 | 24700.9397021 | 24923.9690869 | 223.03 | 0.90 |
| 34 | 27064.3939486 | 27290.5778896 | 226.18 | 0.84 |
| 35 | 29574.5677455 | 29803.4265605 | 228.86 | 0.77 |
| 36 | 32235.8905704 | 32466.9848351 | 231.09 | 0.72 |
| 37 | 35057.6269605 | 35285.7273840 | 228.10 | 0.65 |
| 38 | 38042.8130871 | 38264.1336809 | 221.32 | 0.58 |
| 39 | 41189.2831327 | 41406.6878775 | 217.40 | 0.53 |
| 40 | 44504.2576002 | 44717.8786851 | 213.62 | 0.48 |
| 41 | 47987.0982651 | 48202.1992624 | 215.10 | 0.45 |
| 42 | 51652.6838972 | 51864.1471080 | 211.46 | 0.41 |
| 43 | 55507.1007900 | 55708.2239588 | 201.12 | 0.36 |
| 44 | 59535.0084813 | 59738.9356931 | 203.93 | 0.34 |
| 45 | 63760.7088305 | 63960.7922375 | 200.08 | 0.31 |
| 46 | 68175.4587346 | 68378.3074789 | 202.85 | 0.30 |
| 47 | 72803.4422348 | 72995.9991792 | 192.56 | 0.26 |
| 48 | 77641.5588679 | 77818.3888949 | 176.83 | 0.23 |
| 49 | 82686.1651837 | 82850.0018989 | 163.84 | 0.20 |
| 50 | 87945.2800043 | 88095.3671064 | 150.09 | 0.17 |
| 51 | 93423.5694503 | 93559.0170031 | 135.45 | 0.14 |
| 52 | 99125.8851161 | 99245.4875767 | 119.60 | 0.12 |
| 53 | 105056.3401530 | 105159.3182506 | 102.98 | 0.10 |
| 54 | 111219.4378538 | 111305.0518208 | 85.61 | 0.08 |
| 55 | 117625.4549547 | 117687.2343942 | 61.78 | 0.05 |
| 56 | 124276.8627369 | 124310.4153298 | 33.55 | 0.03 |
| 57 | 131169.6899884 | 131179.1471819 | 9.46 | 0.01 |
| 58 | 138302.0318203 | 138297.9856451 | -4.05 | 0.00 |
| 59 | 145677.7921546 | 145671.4895014 | -6.30 | 0.00 |
| 60 | 153321.8352627 | 153304.2205691 | -17.61 | 0.01 |
| 61 | 161228.7478630 | 161200.7436527 | -28.00 | 0.02 |
| 62 | 169402.7779128 | 169365.6264960 | -37.15 | 0.02 |
| 63 | 177848.3206763 | 177803.4397349 | -44.88 | 0.03 |
| 64 | 186582.1334510 | 186518.7568530 | -63.38 | 0.03 |
| 65 | 195574.5872401 | 195516.1541378 | -58.43 | 0.03 |
| 66 | 204862.9091934 | 204800.2106391 | -62.70 | 0.03 |
| 67 | 214440.8840924 | 214375.5081274 | -65.38 | 0.03 |
| 68 | 224312.9700261 | 224246.6310549 | -66.34 | 0.03 |
| 69 | 234483.4108352 | 234418.1665166 | -65.24 | 0.03 |
| 70 | 244956.5775624 | 244894.7042129 | -61.87 | 0.03 |
| 71 | 255751.3634655 | 255680.8364137 | -70.53 | 0.03 |
| 72 | 266859.0363919 | 266781.1579228 | -77.88 | 0.03 |
| 73 | 278284.5922307 | 278200.2660434 | -84.33 | 0.03 |
| 74 | 290032.4982964 | 289942.7605450 | -89.74 | 0.03 |
| 75 | 302107.0340496 | 302013.2436309 | -93.79 | 0.03 |
| 76 | 314513.8065538 | 314416.3199065 | -97.49 | 0.03 |
| 77 | 327256.1172759 | 327156.5963483 | -99.52 | 0.03 |
| 78 | 340321.8743360 | 340238.6822743 | -83.19 | 0.02 |
| 79 | 353747.2980290 | 353667.1893144 | -80.11 | 0.02 |
| 80 | 367542.1407648 | 367446.7313821 | -95.41 | 0.03 |
| 81 | 381684.6768487 | 381581.9246465 | -102.75 | 0.03 |
| 82 | 396186.2648546 | 396077.3875055 | -108.88 | 0.03 |
| 83 | 411051.7229656 | 410937.7405591 | -113.98 | 0.03 |
| 84 | 426286.0829013 | 426167.6065836 | -118.48 | 0.03 |
| 85 | 441893.3324702 | 441771.6105063 | -121.72 | 0.03 |
| 86 | 457878.0175128 | 457754.3793812 | -123.64 | 0.03 |
| 87 | 474252.9676674 | 474120.5423645 | -132.43 | 0.03 |
| 88 | 491019.4693657 | 490874.7306914 | -144.74 | 0.03 |
| 89 | 508169.9490505 | 508021.5776533 | -148.37 | 0.03 |
| 90 | 525715.2949823 | 525565.7185745 | -149.58 | 0.03 |
| 91 | 543637.9234086 | 543511.7907913 | -126.13 | 0.02 |
| 92 | 561973.8984700 | 561864.4336297 | -109.46 | 0.02 |
| 93 | 580717.6617327 | 580628.2883849 | -89.37 | 0.02 |
| 94 | 599861.4409745 | 599807.9983003 | -53.44 | 0.01 |
| 95 | 619433.5358297 | 619408.2085476 | -25.33 | 0.00 |
| 96 | 639440.7655708 | 639433.5662070 | -7.20 | 0.00 |
| 97 | 659861.8226208 | 659888.7202479 | 26.90 | 0.00 |
| 98 | 680698.4603668 | 680778.3215100 | 79.86 | 0.01 |
| 99 | 701983.6347821 | 702107.0226848 | 123.39 | 0.02 |
| 100 | 723708.1993401 | 723879.4782971 | 171.28 | 0.02 |
| 101 | 745876.4772165 | 746100.3446878 | 223.87 | 0.03 |
| 102 | 768492.9042216 | 768774.2799958 | 281.38 | 0.04 |
| 103 | 791579.9781787 | 791905.9441412 | 325.97 | 0.04 |

**Deriving $\rho(0,Z)$ within and beyond the context of the Bare Coulomb Model (BCM):**

Within the context of the BCM the electron density of an atom is written as:

$$\rho(\vec{r},Z) = \sum_{m=-l}^{l}\sum_{l=0}^{n-1}\sum_{n=1}^{\infty} c_{n,l,m} |\psi_{n,l,m}(\vec{r},Z)|^2 \quad (S1)$$

In this equation $\psi_{n,l,m}(\vec{r},Z)$ are the eigenfunctions of the hydrogen-like atoms ($n,l,m$ are the corresponding quantum numbers) while $c_{n,l,m}$ are the occupation numbers ($\sum_{m=-l}^{l}\sum_{l=0}^{n-1}\sum_{n=1}^{\infty} c_{n,l,m} = N$) of the BCM atom. If one incorporates the radial and angular function then the equations transforms to:

$$\rho(\vec{r},Z) = \sum_{m=-l}^{l}\sum_{l=0}^{n-1}\sum_{n=1}^{\infty} c_{n,l,m} \left| \sqrt{\left(\frac{2Z}{n}\right)^3 \frac{(n-l-1)!}{2n[(n+l)!]}} Exp\left[\frac{-2Zr}{n}\right] \left(\frac{2Zr}{n}\right)^l L_{n-l-1}^{2l+1}\left(\frac{2Zr}{n}\right) Y_{l,m}(\theta,\varphi) \right|^2 \quad (S2)$$

In this equation $L_{n-l-1}^{2l+1}$ and $Y_{l,m}$ are the associated Laguerre polynomials and the spherical harmonics, respectively [S1]. After some mathematical manipulations $\rho(0,Z)$ is derived:

$$\rho(0,Z) = \left(\frac{Z^3}{\pi}\right) \sum_{n=1}^{\infty} c_{n,0,0} \left(\frac{1}{n}\right)^3 \quad (S3)$$

Evidently, only s-orbitals are contributing to the electron density at nucleus. Two "limiting" cases are conceivable for the series namely, $c_{1,0,0} = 2$ and others occupation numbers being zero, and alternatively $c_{n,0,0} = 2$ for all $n$. In the former case only 1s orbital is contributing to the electron density at the nucleus and the known $\rho(0,Z) = (2/\pi)Z^3 \approx 0.637 Z^3$ is derived whereas in the latter case infinite number of the s-orbitals all contributing and the following equation emerges:

$$\rho(0,Z) = \left(\frac{2\xi(3)}{\pi}\right) Z^3 \approx 0.765 Z^3 \quad (S4)$$

where $\xi(s) = \sum_{n=1}^{\infty}\left(\frac{1}{n^s}\right)$ is the Riemann zeta function ($\xi(3) \approx 1.202$) [S1]. For any intermediate situation a coefficient in the narrow range between $\sim 0.637$ and $\sim 0.765$ is conceivable. Evidently, both the coefficient and the exponent of the equation (1) (see the main text), derived from regression procedure namely, $\rho(0,Z) \approx 0.606 Z^{3.039}$, are reproduced properly within the context of the BCM. The following equation is then proposed, heuristically, as a modification of equation (S4):

$$\rho(0,Z) = (1+\alpha)\left(\frac{2\xi(3)}{\pi}\right)(Z + \beta\sqrt{Z})^3 \quad (S5)$$

The coefficient $(1+\alpha)$ is used since upon regression to ab initio $\rho(0,Z)$ dataset only small deviation are observable from the analytical coefficient: $\left(\dfrac{2\xi(3)}{\pi}\right)$, ($\alpha \approx 0.011$ is derived after the regression). Equation (S5) may be written in a more compact form:

$$\rho(0,Z) = a\left(Z - b\sqrt{Z}\right)^3 \tag{S6}$$

where $a = (1+\alpha)\left(\dfrac{2\xi(3)}{\pi}\right)$ and $b = -\beta$. This equation is used for the regression and introduced as equation (2) in the main text.

References:

[S1]   Arfken G (1985) Mathematical Methods for Physicists. Academic Press Inc., San Diego

The ab initio and computed $\rho(0,Z)$ based on equation: $\rho(0,Z)=a(Z-b\sqrt{Z})^3$ ($a \approx 0.773780$, $b \approx 0.220476$) and the absolute error, $\rho(0,Z)-\rho^{ab\ initio}(0,Z)$, and the percentage errors $\left|\rho(0,Z)-\rho^{ab\ initio}(0,Z)\right|/\rho^{ab\ initio}(0,Z)\times 100$. All results are in atomic units ($Z = 2-103$).

| Atomic number | Ab initio | Predicted | Absolute error | Percentage error |
|---|---|---|---|---|
| 2 | 3.5959183 | 3.7229677 | 0.13 | 3.53 |
| 3 | 13.8148199 | 13.8863626 | 0.07 | 0.52 |
| 4 | 35.3877167 | 34.8834328 | -0.50 | 1.43 |
| 5 | 71.9213707 | 70.8403255 | -1.08 | 1.50 |
| 6 | 127.4579724 | 125.9455393 | -1.51 | 1.19 |
| 7 | 205.9683473 | 204.4315269 | -1.54 | 0.75 |
| 8 | 311.6608926 | 310.5639194 | -1.10 | 0.35 |
| 9 | 448.3222768 | 448.6345565 | 0.31 | 0.07 |
| 10 | 619.9220741 | 622.9566680 | 3.03 | 0.49 |
| 11 | 833.7575330 | 837.8613737 | 4.10 | 0.49 |
| 12 | 1093.7178333 | 1097.6950475 | 3.98 | 0.36 |
| 13 | 1402.8456067 | 1406.8172720 | 3.97 | 0.28 |
| 14 | 1765.6069489 | 1769.5992133 | 3.99 | 0.23 |
| 15 | 2186.3141472 | 2190.4223044 | 4.11 | 0.19 |
| 16 | 2669.4696949 | 2673.6771611 | 4.21 | 0.16 |
| 17 | 3219.1903560 | 3223.7626759 | 4.57 | 0.14 |
| 18 | 3839.7818294 | 3845.0852521 | 5.30 | 0.14 |
| 19 | 4538.6539527 | 4542.0581496 | 3.40 | 0.08 |
| 20 | 5319.6071378 | 5319.1009197 | -0.51 | 0.01 |
| 21 | 6182.3043131 | 6180.6389144 | -1.67 | 0.03 |
| 22 | 7133.2302762 | 7131.1028561 | -2.13 | 0.03 |
| 23 | 8176.9553727 | 8174.9284576 | -2.03 | 0.02 |
| 24 | 9313.7337083 | 9316.5560856 | 2.82 | 0.03 |
| 25 | 10560.0774220 | 10560.4304601 | 0.35 | 0.00 |
| 26 | 11908.7041093 | 11911.0003846 | 2.30 | 0.02 |
| 27 | 13367.3818433 | 13372.7185036 | 5.34 | 0.04 |
| 28 | 14940.5685578 | 14950.0410827 | 9.47 | 0.06 |
| 29 | 16625.2578185 | 16647.4278093 | 22.17 | 0.13 |
| 30 | 18447.6758062 | 18469.3416105 | 21.67 | 0.12 |
| 31 | 20397.2023524 | 20420.2484872 | 23.05 | 0.11 |
| 32 | 22480.1179151 | 22504.6173614 | 24.50 | 0.11 |
| 33 | 24700.9397021 | 24726.9199358 | 25.98 | 0.11 |
| 34 | 27064.3939486 | 27091.6305647 | 27.24 | 0.10 |
| 35 | 29574.5677455 | 29603.2261342 | 28.66 | 0.10 |
| 36 | 32235.8905704 | 32266.1859509 | 30.30 | 0.09 |
| 37 | 35057.6269605 | 35084.9916395 | 27.36 | 0.08 |
| 38 | 38042.8130871 | 38064.1270465 | 21.31 | 0.06 |
| 39 | 41189.2831327 | 41208.0781505 | 18.80 | 0.05 |
| 40 | 44504.2576002 | 44521.3329790 | 17.08 | 0.04 |
| 41 | 47987.0982651 | 48008.3815298 | 21.28 | 0.04 |
| 42 | 51652.6838972 | 51673.7156974 | 21.03 | 0.04 |
| 43 | 55507.1007900 | 55521.8292045 | 14.73 | 0.03 |
| 44 | 59535.0084813 | 59557.2175367 | 22.21 | 0.04 |
| 45 | 63760.7088305 | 63784.3778814 | 23.67 | 0.04 |
| 46 | 68175.4587346 | 68207.8090701 | 32.35 | 0.05 |
| 47 | 72803.4422348 | 72832.0115240 | 28.57 | 0.04 |
| 48 | 77641.5588679 | 77661.4872022 | 19.93 | 0.03 |
| 49 | 82686.1651837 | 82700.7395533 | 14.57 | 0.02 |
| 50 | 87945.2800043 | 87954.2734686 | 8.99 | 0.01 |
| 51 | 93423.5694503 | 93426.5952383 | 3.03 | 0.00 |
| 52 | 99125.8851161 | 99122.2125099 | -3.67 | 0.00 |
| 53 | 105056.3401530 | 105045.6342484 | -10.71 | 0.01 |
| 54 | 111219.4378538 | 111201.3706984 | -18.07 | 0.02 |
| 55 | 117625.4549547 | 117593.9333481 | -31.52 | 0.03 |
| 56 | 124276.8627369 | 124227.8348947 | -49.03 | 0.04 |
| 57 | 131169.6899884 | 131107.5892120 | -62.10 | 0.05 |
| 58 | 138302.0318203 | 138237.7113188 | -64.32 | 0.05 |
| 59 | 145677.7921546 | 145622.7173486 | -55.07 | 0.04 |
| 60 | 153321.8352627 | 153267.1245215 | -54.71 | 0.04 |
| 61 | 161228.7478630 | 161175.4511162 | -53.30 | 0.03 |
| 62 | 169402.7779128 | 169352.2164438 | -50.56 | 0.03 |
| 63 | 177848.3206763 | 177801.9408225 | -46.38 | 0.03 |
| 64 | 186582.1334510 | 186529.1455535 | -52.99 | 0.03 |
| 65 | 195574.5872401 | 195538.3528973 | -36.23 | 0.02 |
| 66 | 204862.9091934 | 204834.0860518 | -28.82 | 0.01 |
| 67 | 214440.8840924 | 214420.8691303 | -20.01 | 0.01 |
| 68 | 224312.9700261 | 224303.2271409 | -9.74 | 0.00 |
| 69 | 234483.4108352 | 234485.6859669 | 2.28 | 0.00 |
| 70 | 244956.5775624 | 244972.7723470 | 16.19 | 0.01 |
| 71 | 255751.3634655 | 255769.0138570 | 17.65 | 0.01 |
| 72 | 266859.0363919 | 266878.9388922 | 19.90 | 0.01 |
| 73 | 278284.5922307 | 278307.0766498 | 22.48 | 0.01 |
| 74 | 290032.4982964 | 290057.9571122 | 25.46 | 0.01 |
| 75 | 302107.0340496 | 302136.1110312 | 29.08 | 0.01 |
| 76 | 314513.8065538 | 314546.0699124 | 32.26 | 0.01 |
| 77 | 327256.1172759 | 327292.3660000 | 36.25 | 0.01 |
| 78 | 340321.8743360 | 340379.5322624 | 57.66 | 0.02 |
| 79 | 353747.2980290 | 353812.1023781 | 64.80 | 0.02 |
| 80 | 367542.1407648 | 367594.6107222 | 52.47 | 0.01 |
| 81 | 381684.6768487 | 381731.5923530 | 46.92 | 0.01 |
| 82  | 396186.2648546 | 396227.5829993 | 41.32  | 0.01 |
| 83  | 411051.7229656 | 411087.1190483 | 35.40  | 0.01 |
| 84  | 426286.0829013 | 426314.7375330 | 28.65  | 0.01 |
| 85  | 441893.3324702 | 441914.9761212 | 21.64  | 0.00 |
| 86  | 457878.0175128 | 457892.3731038 | 14.36  | 0.00 |
| 87  | 474252.9676674 | 474251.4673841 | -1.50  | 0.00 |
| 88  | 491019.4693657 | 490996.7984668 | -22.67 | 0.00 |
| 89  | 508169.9490505 | 508132.9064479 | -37.04 | 0.01 |
| 90  | 525715.2949823 | 525664.3320047 | -50.96 | 0.01 |
| 91  | 543637.9234086 | 543595.6163860 | -42.31 | 0.01 |
| 92  | 561973.8984700 | 561931.3014023 | -42.60 | 0.01 |
| 93  | 580717.6617327 | 580675.9294171 | -41.73 | 0.01 |
| 94  | 599861.4409745 | 599834.0433374 | -27.40 | 0.00 |
| 95  | 619433.5358297 | 619410.1866053 | -23.35 | 0.00 |
| 96  | 639440.7655708 | 639408.9031893 | -31.86 | 0.00 |
| 97  | 659861.8226208 | 659834.7375760 | -27.09 | 0.00 |
| 98  | 680698.4603668 | 680692.2347623 | -6.23  | 0.00 |
| 99  | 701983.6347821 | 701985.9402471 | 2.31   | 0.00 |
| 100 | 723708.1993401 | 723720.4000241 | 12.20  | 0.00 |
| 101 | 745876.4772165 | 745900.1605738 | 23.68  | 0.00 |
| 102 | 768492.9042216 | 768529.7688567 | 36.86  | 0.00 |
| 103 | 791579.9781787 | 791613.7723060 | 33.79  | 0.00 |